\documentclass[11pt,a4]{article}

\usepackage{latexsym}
\usepackage[dvips]{graphicx}
\usepackage{amsmath}
\usepackage{amssymb}

\usepackage{times}
\usepackage{bm}

\usepackage{enumerate}
\usepackage{amsfonts}
\usepackage{amsthm}
\usepackage{mathrsfs}
\usepackage{ascmac}

\newcommand{\half}{\frac{1}{2} }

\newcommand{\norm}[1]{\left\| #1  \right\| }

\newcommand{\real}{\mathbf{R}}
\newcommand{\complex}{\mathbf{C}}

\newcommand{\Hilbert}{\mathcal{H} }

\newcommand{\state}{\mathcal{S}(\mathcal{H})}

\newcommand{\model}{\mathcal{M}}

\newcommand{\povm}{\mathcal{P}o}

\renewcommand{\cal}[1]{\mathcal{#1}}

\newcommand{\etal}{\textit{et al.}}

\newcommand{\DS}{\displaystyle}

\newcommand{\dtheta}{\mathrm{d}\theta}

\newcommand{\Tr}{\mathrm{Tr}}

\newcommand{\migi}{\rightarrow}

\newcommand{\Mre}{\mathbf{M}}

\newcommand{\rmd}{\mathrm{d}}

\newcommand{\ran}{\mathrm{ran}}

\theoremstyle{definition}
\newtheorem{theorem}{Theorem}
\newtheorem{corollary}[theorem]{Corollary}
\newtheorem{proposition}{Proposition}
\newtheorem{lemma}{Lemma}
\newtheorem{remark}{Remark}
\newtheorem{definition}{Definition}
\newtheorem{example}{Example}

\newcommand{\ptheta}{\cal{P}(\Theta) }

\begin{document}

\setlength{\baselineskip}{20pt}

\title{Quantum Minimax Theorem}

\author{Fuyuhiko TANAKA}


\maketitle

\begin{abstract}
Recently, many fundamental and important results in statistical decision theory have been extended to the quantum system.
Quantum Hunt-Stein theorem and quantum locally asymptotic normality are typical successful examples.
In the present paper, we show quantum minimax theorem, which is also an extension of a well-known result, minimax theorem
 in statistical decision theory, first shown by Wald and generalized by Le~Cam.
Our assertions hold for every closed convex set of measurements and for general parametric models of density operator.
On the other hand, Bayesian analysis based on least favorable priors has been widely used in classical statistics
 and is expected to play a crucial role in quantum statistics.
According to this trend, we also show the existence of least favorable priors, which seems to be new even in classical statistics.
\end{abstract}


%
\section{Introduction}
\label{intro}






Quantum statistical inference is the inference on a quantum system from relatively small amount of measurement data.
It covers also precise analysis of statistical error~\cite{Sugiyama2013}, exploration of optimal measurements to extract information~\cite{MassarPopescu1995}, development of efficient numerical computation~\cite{DiGuglielmo2009}. 
With the rapid development of experimental techniques, there has been much work on quantum statistical inference~\cite{Paris2004}, which is now applied to quantum tomography~\cite{Artiles2005,Butucea2007}, validation of entanglement~\cite{Haseler2008}, and quantum benchmarks~\cite{Haseler2010,Namiki2008b}.
In particular, many fundamental and important results in statistical decision theory~\cite{Wald1950} have been extended to the quantum system.
Theoretical framework was originally established by Holevo~\cite{Holevo1973a,Holevo1976,Holevo}.
Quantum Hunt-Stein theorem~\cite{Holevo,Ozawa1980,Bogomolov1982} and quantum locally asymptotic normality~\cite{Guta2007,Kahn2009} are typical successful examples.


On the other hand, gradual increase in the size of quantum systems in an experiment causes different kinds of statistical problems. 
As Gross~\etal~\cite{Gross2010} pointed out, 
even in a quantum state of $8$ ions, a maximum-likelihood estimate 
required hundreds of thousands of measurements and weeks of post-processing.
They proposed a statistical method based on compressed sensing and 
show that it has much better performance than ordinary estimating methods if the prepared quantum state is nearly pure.
Blume-Kohout also pointed out this practical issue and explains how the Bayesian estimate is useful~\cite{Blume2010}.
Another promising approach is to choose a suitable parametric model that has a limited number of parameters as in Yin and van~Enk~\cite{YinvanEnk2011b}.
Constructed models heavily depend on the experimental setup and may lack symmetry.
However, they would be more useful than group acting models with many parameters to be specified.

%

These efforts in the second paragraph seem to have no relation to fundamental, important achievement in quantum statistical decision theory
 described in the latter of the first paragraph.
Quantum Hunt-Stein theorem is a mathematically beautiful result and gives an optimal POVM in a group acting model
 (e.g., quantum hypothesis testing of quantum Gaussian states in Kumagai and Hayashi~\cite{Kumagai2013}.) 
However, as in classical statistics, experimenters may have some knowledge on a prepared system, e.g., some mean parameters are positive.
Quantum locally asymptotic normality~\cite{Guta2007,Kahn2009} would tell us an optimal POVM in an asymptotic sense for any parametric model of density operators.
However, there are some tricks.
First, if we expect the large number of data for the prepared system and implicitly assume high performance of data processing, 
 then more na\"{i}ve method may be sufficient for this situation.
Second, more importantly and different from classical statistics, a mathematically optimal POVM may not be easy to implement in the experimental system.


In the present paper, we show quantum minimax theorem, which is also an extension of a well-known result, minimax theorem
 in statistical decision theory, first shown by Wald~\cite{Wald1950} and generalized by Le~Cam~\cite{LeCam1964}. 
However, not only from purely mathematical or theoretical motivation, but also from more practical motivation
 we deal with the theorem.
We emphasize three points.
Our result is
\begin{enumerate}[(i)]
\item available for any parametric model (i.e., no or less group symmetry) 
\item under finite sample assumption (i.e., no asymptotics)
\item also effective when we restrict POVMs to a smaller class (i.e., that may depend on the experimenter's preference)
\end{enumerate}

Technically speaking, meaningful results in statistics without resort to asymptotic methods and group symmetry are often hard to prove.
We have used many tools in functional analysis.



Since our theorem holds under very general assumptions, it could be potentially applied to a broad range of quantum statistical problems.
In order to show how our theorem works, let us take a simplified example in quantum state discrimination, which is also regarded as a discretized version of quantum state estimation.
We also explain some concepts in statistical decision theory.
(For quantum state discrimination, see e.g., Eldar \etal~\cite{Eldar2004b} and references therein. For the meaning of each concept in statistical decision theory, see, e.g., Ferguson~\cite{Ferguson}.)
Suppose that Alice has three quantum states
\begin{align*}
& \rho_{1} = \begin{pmatrix}
	1 & 0 & 0 \\
	0 & 0 & 0 \\
	0 & 0 & 0 
 \end{pmatrix}, \rho_{2} = \begin{pmatrix}
	1/2 & 1/2 & 0 \\
	1/2& 1/2  & 0 \\
	0 & 0& 0 
 \end{pmatrix}, \rho_{3} =  \begin{pmatrix}
	0 & 0 & 0 \\
	0 & 0 & 0 \\
	0 & 0 & 1 
\end{pmatrix}, 
\end{align*}
 and randomly chooses one state and sends it to Bob.
Bob prepares a POVM to determine which the received quantum state really is.
The POVM $\Mre$ is given by $M_{1}, M_{2}, M_{3}$, where each element is a three-dimensional positive semidefinite  matrix and
 $M_{1} +M_{2} + M_{3} = I$.
When Alice sends $i$-th state, the probability that Bob obtains the outcome $j$ is given by
\[
p_{\Mre}(j | i) = \Tr \rho_{i} M_{j}.
\]
In this setting, we will find a good POVM.
In order to discuss in a quantitative way, we set Bob's loss in the following manner: 
Bob gets zero if his guess is  correct and gets one if his guess is wrong.
Using Kronecker's delta, the loss is given by a function of pair $(i, j)$, 
\[
 w(i, j) = 1- \delta_{ij},\ i, j =1,2,3.
\]
Then, the expected loss for Bob conditional to Alice's choice is given by
\[
R_{\Mre}(i) := \sum_{j=1}^{3} w(i, j) p_{\Mre}(j | i),
\]
 which is called a \textit{risk function}.
For each $i$, smaller risk is better.
Since $\rho_{1}$ and $\rho_{2}$ are nonorthogonal to each other, there is no POVM that achieves 
the minimum risk (i.e., zero) for every $i$.
As is shown later, in statistical decision theory, we consider two optimality criteria.

Suppose that Bob has some knowledge on Alice's choice and it is written as a probability distribution, $\pi (1) + \pi(2) + \pi(3) =1$, 
which is called a \textit{prior distribution} or shortly \textit{prior}.
Then he might consider the average risk,
\[
 r_{\Mre} (\pi ) := \sum_{i=1}^{3} R_{\Mre} (i) \pi(i),
\]
which is a scalar function of Bob's POVM $\Mre $.
In this setting, there exists a minimizer, which is called a \textit{Bayesian POVM} (a \textit{Bayes POVM}) with respect to $\pi$.
On the other hand, if Bob has no knowledge on Alice's choice, then he may consider the worst case risk
\[
 r_{\Mre}^{SUP}  := \sup_{i} R_{\Mre} (i), 
\]
 which is again a scalar function of $\Mre $.
There exists a minimizer in this case and it is called a \textit{minimax POVM}.

Bayes POVM and minimax POVM are defined separately and derived from independent optimality criterion.
Quantum minimax theorem gives a deep relation between them.
More explicitly, the theorem asserts,
\begin{equation} \label{disc:minimax} 
 \inf_{\Mre} r_{\Mre}^{SUP} = \sup_{\pi} \inf_{\Mre} r_{\Mre} (\pi ).
\end{equation}
Roughly speaking, quantum minimax theorem states that a minimax POVM is given by a Bayes POVM with respect to a certain prior, which is called a \textit{least favorable prior}.
Intuitively speaking, Alice's choice based on a least favorable prior corresponds to the worst case to Bob.

In the above example, a least favorable prior is unique and given by $\pi_{LF} (1) = \pi_{LF}(2) =1/2, \pi_{LF}(3)=0$.
How do we obtain the prior?
Observe that Bob can distinguish the state $\rho_{3}$ from the other states $\rho_{1}$ and $\rho_{2}$ perfectly.
If Alice sends $\rho_{3}$ to Bob with a positive probability, then he detects $\rho_{3}$ with no error.
Thus, the best strategy for Alice to beat Bob is to neglect $\rho_{3}$ and to choose either $\rho_{1}$ or $\rho_{2}$
 with equal probability.



Bayesian POVM with respect to $\pi_{LF}$ is given by 
\begin{align*}
& M_{1} =\half  \begin{pmatrix}
	1 +1/\sqrt{2} & -1/\sqrt{2} & \phantom{0} 0 \\
	-1/\sqrt{2} & 1-1/\sqrt{2} & \phantom{0}0 \\
	0 & 0 & \phantom{0}0 
\end{pmatrix}, 
& M_{2} = \half  \begin{pmatrix}
	1 -1/\sqrt{2} & 1/\sqrt{2} & \phantom{0}0 \\
	1/\sqrt{2} & 1+1/\sqrt{2} & \phantom{0}0 \\
	0 & 0 & \phantom{0}0 
\end{pmatrix}, \\
& M_{3} =  \begin{pmatrix}
	0\phantom{0} & 0\phantom{0} & 0\phantom{0} \\
	0\phantom{0} & 0\phantom{0} & 0\phantom{0} \\
	0\phantom{0} & 0\phantom{0} & 1\phantom{0} 
\end{pmatrix}, 
\end{align*}
 which is indeed shown to be minimax.

We emphasize that the whole set of quantum states $\{ \rho_{1}, \rho_{2}, \rho_{3} \}$ have no symmetry.
In spite of this, we successfully obtain a good POVM.
The above POVM is both minimax and Bayes with respect to a prior $\pi_{LF}$.
Thus Eq.~\eqref{disc:minimax} assures the existence of a good POVM.
(Statistically speaking, the POVM is called \textit{admissible}. 
In the present paper, we do not discuss admissibility and related concepts.
They will be presented for another more statistical journal.)
If Eq.~\eqref{disc:minimax} does not hold in a problem, both minimax POVM and Bayes POVM lack theoretical justification.

The above claims are not restricted to the case where a finite set of density operators are given.
As is shown later, our result covers not only quantum state discrimination but also quantum estimation and quantum state tomography.
For any lower-semicontinuous loss function, which covers usual loss functions in quantum statistical inference, 
 we show that any minimax POVM is also Bayes with respect to $\pi_{LF}$ under some assumptions like compactness.
It is a direct consequence of quantum minimax theorem (Theorem~\ref{theo:minimax2}), which is a general version of Eq.~\eqref{disc:minimax}.
In addition, if we restrict POVMs to a smaller class of POVMs, e.g., PVMs and its randomization, or separable POVMs over a composite system,
 still our assertion holds.
Thus, our result possibly gives a guideline to many experimenters in quantum physics.
For example, the performance of the Bayes POVM $\Mre_{\pi}$ with respect to $\pi$ (a minimizer in a given class of POVMs)
 is evaluated by the following difference,
\begin{equation} \label{value:diff}
 \sup_{\theta \in \Theta} R_{\Mre_{\pi} } (\theta ) -  r_{\Mre_{\pi}}(\pi ).
\end{equation}
This term is necessarily nonnegative by definition.
If the term~\eqref{value:diff} is close to zero, it implies that $\Mre_{\pi}$ is nearly optimal among a given class of POVMs 
 and $\pi$ is nearly least favorable.









Here we mention related works in classical statistics.
The concept of least favorable priors plays a crucial role in (objective) Bayesian statistics.
For example, Bayesian analysis based on reference priors has been widely recognized among 
statisticians~\cite{Bernardo1979,BernardoSmith,Bernardo2005} and the reference prior is formally defined as a least favorable prior.
Recent and significant development in this direction is the latent information prior by Komaki~\cite{Komaki2011}.
The latent information prior is also a least favorable prior.
Previous results on least favorable priors suggest that ``least favorable" or ``least information" does not necessarily agree with our intuition of no information.
It is also interesting to see least favorable priors in quantum pure states models, which consist of
 pure states~\cite{FT2012b}.
Even if three nonorthogonal pure states are given, complete ignorance is no longer described by the uniform prior (i.e., the same weight to each state vector).



Apart from Introduction, we focus on mathematical aspects of the entire scenario,
 that is, proof of quantum minimax theorem and its corollary.
We also show the existence of least favorable priors, which seems to be new in classical statistics.
If we would have followed Le~Cam's path in a mathematically straightforward way without any statistical consideration, 
we would not have been able to show the existence in a nontrivial way.
Applications in quantum state tomography with more practical issues and investigations from a more statistical viewpoint will be presented separately for another occasion.

In the next section, we briefly review Le~Cam's fundamental result~\cite{LeCam1964}, which is partially used in the proof of quantum minimax theorem.
In Section 3, we introduce quantum statistical decision theory following Holevo~\cite{Holevo1976}. 
We adopt the Hilbert space formalism to describe quantum systems instead of algebraic one because recent many works in quantum statistical inference are described in this way.
Some statistical concepts are from statistical decision theory by Wald~\cite{Wald1950}. We often see them in an advanced textbook in mathematical statistics~\cite{Ferguson}.
In Section 4, we show the quantum minimax theorem.
In Section 5, we introduce least favorable priors and show the existence theorem.
Concluding remarks are given in the last section.
Proof of Lemma 2, which is crucial in our argument, is a bit tedious and presented in Appendix.


\section{Preliminary}
\label{sec:2}


First we briefly review Le~Cam's fundamental result~\cite{LeCam1964}.
Here, we follow the standard textbook by Strasser~\cite{Strasser1985} rather than Le~Cam's original article.
Readers who are not familiar to functional analysis may skip technical assumptions and details.

Let  $\Theta $ be a locally compact topological space and $C(\Theta )$ be the set of all continuous functions on $\Theta$.
The space $C (\Theta)$ is topologized with the topology of uniform convergence on compact sets.
Let  $\ptheta $ be the space of all probability measures on Borel sets $\beta (\Theta )$ with compact support.

\begin{definition} 
Let $\cal{M} \subseteq C(\Theta )$ be an arbitrary set.
The lower envelope of $\model$ is the function
\[
 \pi \mapsto \psi_{\model} (\pi ) := \inf_{g \in \model} \int g(\theta ) \pi ( \dtheta ),\  \pi \in \cal{P} (\Theta ).
\]
\end{definition}
\begin{definition}
Let $\cal{M} \subseteq C(\Theta )$ be an arbitrary set. Then
\[
\alpha (\model) :=\bigcup_{f \in \model} \{ g \in C(\Theta ):\ f \leq g \}.
\]
\end{definition}

\begin{lemma}
\label{lem:1}
For every $\model \subseteq C(\Theta )$ and $\pi \in \ptheta$,
\begin{equation} \label{eq:alpham}
\psi_{\model}(\pi) = \psi_{\overline{\alpha (\model ) }  } (\pi )
\end{equation}
holds.
In particular, when $\model $ is compact, 
\begin{equation} \label{eq:alpham2}
\psi_{\model}(\pi) = \psi_{ \alpha (\model)   } (\pi ).
\end{equation}
\end{lemma}

\proof
By definition, it is easily seen that \eqref{eq:alpham} holds.
When $\model$ is compact, $\alpha (\model )$ is closed.
Thus $\overline{\alpha (\model )} = \alpha (\model )$ holds.
\qed

\begin{definition}
A set $\cal{M} \subseteq C(\Theta )$ is called \textit{subconvex} if for $f_{1} \in \cal{M}, f_{2} \in \cal{M}$ and $ \alpha \in (0,1)$
 there is  $f_{3} \in \cal{M}$ such that 
$ f_{3} \leq \alpha f_{1} + (1- \alpha ) f_{2}$. 
\end{definition}
Every convex set is subconvex. If $\model $ is subconvex then $\alpha (\model ) $ is convex. 



The following proposition is due to Le~Cam.

\begin{proposition} \label{pro:LeCam}~\cite{LeCam1964}
Let  $\cal{M}_{1}, \cal{M}_{2} \subseteq C(\Theta )$.
Assume that $\cal{M}_{2}$ is subconvex.
Then the following assertions are equivalent: 
\begin{enumerate}[(i)]
\item $f \in \cal{M}_{1}$ $\exists g \in \overline{\alpha (\cal{M}_{2})}$ such that $g (\theta ) \leq f(\theta ),\ \forall \theta \in \Theta $.
\item 
$ \psi_{\cal{M}_{2}} (\pi ) \leq \psi_{\cal{M}_{1}} (\pi ),\ \forall \pi \in \ptheta $.
\end{enumerate}
\end{proposition}

Originally Le~Cam introduced the discrete topology in $\Theta $ in order to show the general kind of minimax theorem and some consequences in statistical decision theory~\cite{LeCam1964}.
In this case, any function on $\Theta$ is regarded as a continuous function.
Certainly technical difficulties are avoided when we use the discrete topology
 but we often consider continuous parameter spaces with our common sense of the continuity. 
Thus we do not restrict the topology of $\Theta$.
Instead, in Section~\ref{sec:3}, we impose an additional assumption on a parametric model of density operators $\{ \rho (\theta) \}$.
This assumption is more practical because usual parametric models satisfy it.

\section{Quantum statistical decision theory}
\label{sec:3}

Next we focus on our problem.
Let $\Hilbert $ be a Hilbert space and $L_{1}(\Hilbert)$ denote the set of all trace-class operators. 

\subsection{Quantum statistical models}


First of all, we define quantum statistical models.
Holevo~\cite{Holevo1976} gives a condition where an operator-valued integral is properly defined.
We see this condition and later we assume it for quantum statistical models.
Indeed the condition is essentially used to show Lemma~\ref{lemma:myown}.

For notational convenience, we define
$K_{\delta} :=  \{ (\theta, \eta) \in K \times K:\ d(\theta, \eta) < \delta \}$,
for every positive $\delta$ and every compact set $K$.
Now let us introduce a norm-like quantity.
For a trace class operator-valued function $T: \Theta \to L_{1}(\Hilbert)$, we define
\begin{align*}
&\omega_{T} (K_{\delta} ) := \inf \{  \norm{X}_{1}:\ -X \leq T (\theta )- T (\eta ) \leq X,\ \forall (\theta, \eta) \in K_{\delta } \}
\end{align*}
for a positive $\delta $ and a compact set $K$, where $A \geq B$ if and only if an operator $A-B$ is positive.\\

\begin{definition}
A function $T$ from $\Theta $ to $L_{1}(\Hilbert)$ is said to be \textit{regular} if it satisfies the following condition, 
\[
 \lim_{\delta \to 0} \omega_{T} (K_{\delta}) =0,\ 
\]
for every compact set $K$. 
\end{definition}

The regularity is slightly stronger than the uniform continuity with respect to the trace norm.
It is easily seen that a regular function is uniformly continuous on every compact set with respect to the trace norm.
Generally the converse does not hold in an infinite-dimensional Hilbert space.
As far as the author knows, such cases are exceptional.

Roughly speaking, for a regular trace class operator-valued function $T$, the operator-valued integral 
\[
 \int_{\Theta} f(\theta ) T (\theta ) \pi (d \theta )
\]
is properly defined for every continuous function $f \in C(\Theta )$ and every probability measure $\pi \in \ptheta $~\cite{Holevo1976}.

\begin{definition}
A function $\rho: \Theta \to L_{1}(\Hilbert)$ is called a \textit{quantum statistical model}
if it satisfies 
$\rho (\theta ) \geq 0, \  \Tr \rho (\theta ) =1$.
\end{definition}

We assume the following conditions:
\begin{enumerate}[(i)]
\item Identifiability, $\rho (\theta_{1}) \neq \rho (\theta_{2})$ if $\theta_{1} \neq \theta_{2}$.
\item Regularity, 
\[
 \lim_{\delta \to 0} \omega_{\rho} (K_{\delta}) =0,\ 
\]
for every compact set $K$. 
\end{enumerate}

\begin{remark}
Some readers might be familiar to the notation like $ \{ \rho (\theta) :\ \theta \in \Theta \}$.
Our definition agrees with the usual definition of quantum statistical models without emphasis on a regularity.
\end{remark}


Let us take a simple example in order to help readers understand these concepts.

\begin{example}
Let us consider two-dimensional Hilbert spaces ($\dim \Hilbert =2$).
In physics, two-level system in atom, spin part of elementary particles, photon polarization are described by a suitable density operator in $\complex^{2}$. 
We take one specific parametric model of a density operator as an example.
\begin{align*}
\rho (\theta ) &= \half 
\begin{pmatrix}
	1 & \theta  \\
	\theta  & 1 
\end{pmatrix},
\end{align*}
 where $\theta $ is a real parameter.
From positivity of $\rho (\theta)$, we obtain $ -1 \leq \theta \leq 1$.
\end{example}

\subsection{Loss functions}

Let $U$ be a locally compact topological space.
The space $U$ plays the role of the decision space.

\begin{definition}
Let $w: \Theta \times U \migi \real\cup \{ + \infty\} $ be a lower semicontinuous function.
We call $w$ a \textit{loss function} if it is bounded from below, $w(\theta, u) > -M > -\infty, \forall \theta, \forall u$ for a constant $M$.
\end{definition}

For simplicity, we assume that $w(\theta, u) \geq 0$.

\begin{definition}
For a quantum statistical model $\rho (\theta )$, a decision space $U$ and a loss function $w(\theta, u)$,
 we call the triplet $(\rho, U, w) $ a \textit{quantum statistical decision problem}. \\
\end{definition}

Classical statistical inference is usually formulated as a statistical decision problem.
Likewise, quantum statistical inference including quantum estimation, quantum state discrimination is formulated as a quantum statistical decision problem. 
(See, e.g., references in Kumagai and Hayashi~\cite{Kumagai2013} for recent works in this direction, although they deal with non-Bayesian hypothesis testing.)
Interestingly enough, quantum state cloning~\cite{Bruss1998,Fan2002} and benchmark fidelity~\cite{Hammerer2005,Haseler2008,Haseler2010,Namiki2008b}, which are purely physical topics, are also described in the framework (See, e.g., Tanaka~\cite{FT2012c} for the relation between quantum benchmark and quantum estimation of pure states).

In classical statistics, the observation $x$ is a random variable and
 its distribution is given as a member of a parametric model $\{ p(x| \theta )\}$.
We only have to specify a decision function
 $\delta: x \mapsto \delta (x) $.
In the quantum setting, however, we have to specify a measurement over the quantum system, which is described 
by a positive-operator-valued measure and then the distribution of the observation $x$ is determined by the measurement
 and the density operator.



\subsection{POVMs and weak topology}
\label{povms}


We give a mathematical description of measurement according to Holevo~\cite{Holevo1976}.
For later convenience, we adopt the Borel sets as a $\sigma$-algebra rather than the Baire sets.
Since the Baire sets are included in the Borel sets on a locally compact space, our definition becomes more simple than that in Holevo. (See, e.g., Royden section 13-1~\cite{Royden} for definitions of Borel sets and Baire sets.)

Let $\cal{A}(U)$ be the $\sigma$-algebra of Borel sets and  $\Mre =\{ M (B )\}$ be a positive operator-valued function on $\cal{A}(U)$ satisfying the following conditions:   
\begin{enumerate}
\item [(i)]  $ M(B) \geq 0,\ B \in \cal{A}(U)$,
\item [(ii)] If $ B = \cup_{j}B_{j}$, $B_{j} \in \cal{A}(U)$ and $B_{i} \cap B_{j} = \emptyset$  for $i \neq j$, then 
 $M(B) = \sum_{j} M(B_{j})$, where the series converges in the weak topology in $\cal{B} (\Hilbert) $. 
\item [(iii)] $ M(U) = I$.
\end{enumerate}
We call $\Mre $ a \textit{measurement} or a \textit{POVM}. 
For any density operator $\rho $, a POVM yields a probability distribution of measurement outcome $u$
 by the Born rule, 
\[
  \mu_{\rho, \Mre}(B) := \Tr \rho M (B),\ B \in \cal{A}(U).
\]
Thus, the integral on $U$ is properly defined with respect to this measure.

Let $\povm (U)$ be the class of all POVMs. (We often omit the decision space $U$.)
Next, we introduce a topology according to Holevo~\cite{Holevo1976} by defining open neighborhoods of a POVM $\Mre \in \povm (U)$. 
For every $\epsilon > 0$ and every positive trace-class operator $T$, let us define
 a finite measure 
\[
\mu_{T, \Mre} (A) := \Tr T M (A),\ A \in \cal{A}(U).
\]
Then the integral of $f \in C(U)$ with respect to the measure 
is defined by
\[
 \mu_{T, \Mre} [f] := \int_{U} f(u) \mu_{T, \Mre}  (\rmd u).
\]

\begin{definition}
\label{def:topology}
We define the system of neighborhoods of $\Mre$, 
\[
\cal{U}(\Mre; \epsilon, T, f)
:= \bigl\{  \Mre' \in \povm (U) :\ |  \mu_{T, \Mre} [f]  -\mu_{T, \Mre'} [f]  | < \epsilon   \bigr\}.
\]
A topology on $\povm (U)$ is generated by finite product of such neighborhoods.
\end{definition}

The following Lemma is due to Holevo (Chapter II, Section 4, p. 53 in Holevo~\cite{Holevo1976}).
\begin{lemma}
If $U$ is compact, then $\povm (U)$ is compact with respect to the above topology.
In particular, every closed subset of POVM is also compact.
\end{lemma}



\begin{example} (continued)
Let us consider a projective measurement described by
\begin{align*}
E_{1} &:=
	\half 
	\begin{pmatrix}
		1 & 1 \\
		1 & 1 \\
	\end{pmatrix}, &
E_{-1} &:=
	\half 
	\begin{pmatrix}
		1 & -1 \\
		-1 & 1 \\
	\end{pmatrix}.
\end{align*}
The set of measurement outcome here is $\{ 1,-1 \}$.
If we repeat the same measurement twice, then the quantum statistical model becomes $\{ \rho (\theta)^{\otimes 2} \}$.
The number that we obtain the outcome $1$, $n_{1}$, is distributed according to the binomial distribution.
When we consider parameter estimation, our decision space is $U = \Theta = [0,1]$.
One typical estimate of $\theta$ is $\delta(n_{1}) = 2(n_{1}/2) - 1$ because the expectation of $\delta (n_{1})$
 is equal to $\theta$. (In statistics, such an estimate is called an unbiased estimate.)
Then, our whole estimating process is described by the following POVM:
\begin{align*}
F_{1} := E_{1} \otimes E_{1}, \ F_{1/2} := E_{1} \otimes E_{-1} + E_{-1} \otimes E_{1}, F_{0} := E_{-1} \otimes E_{-1}.
\end{align*}
Our estimate $u$ is distributed according to 
\[
 u \sim \Tr \rho (\theta)^{\otimes 2} F_{u},
\] which agrees with the distribution of $\delta (n_{1})$.
The average squared error is given by
\[
 \sum_{u=0, 1/2, 1} (u - \theta )^{2}  \Tr \rho (\theta)^{\otimes 2} F_{u}.
\]
\end{example}


\subsection{Risk functions and optimality}
\label{risk}

In statistical decision theory, the average of the loss function plays the fundamental role.
\begin{definition}
\label{def:risk}
For a given quantum statistical decision problem, 
\[
R_{\Mre} (\theta ) := \int_{U} w(\theta, u) \Tr \rho (\theta ) M (du)
\]
is called a \textit{risk function with respect to a POVM $\Mre$}.
\end{definition}

In terms of the risk function (of course, smaller is better),  we are able to discuss which POVM is good or bad, what is the best POVM among a certain class of POVMs like group covariant measurements, or experimentally feasible measurements.
Apart from exceptional cases, almighty POVM does not exist in a given decision problem.

We have at least two optimality criteria, one is Bayesian and the other is minimaxity.
If we are given a probability distribution $\pi \in \ptheta $, then we are able to consider 
the minimum of the average risk,
\[
 r_{\Mre} := \int_{\Theta} R_{\Mre}(\theta)  \pi (d \theta ).
\]
The minimizer is called a \textit{Bayesian POVM (Bayes POVM) with respect to} $\pi$ if it exists.
Usually a Bayesian POVM depends on a distribution $\pi $.

On the other hand, we may consider the minimization of the worst case risk 
$\sup_{\theta \in \Theta} R_{\Mre} (\theta )$.
The minimizer is called a \textit{minimax POVM} if it exists.


\section{Quantum minimax theorem}
\label{qminimax}

Now we are ready to mention our main result. We show the following equality under additional assumptions.
\[
\inf_{\Mre \in \povm } \sup_{\theta \in \Theta}
 R_{\Mre} (\theta )
 = \sup_{\pi \in \ptheta} \inf_{\Mre \in \povm} \int R_{\Mre } (\theta ) \pi (\dtheta ).
\]
This equality clarifies the deep relation between minimax POVMs and Bayesian POVMs.
The classical counterpart was shown decades ago~\cite{Wald1950,LeCam1964} and was called minimax theorem in statistical decision theory.
Thus we call the above equality \textit{quantum minimax theorem} shortly.

From now on we assume that both $\Theta$ and $U$ is compact metric space respectively.
Let $C(\Theta )$ be the set of all continuous functions on $\Theta $
 and topologized with the supremum norm. 
We also assume that $w (\theta, u)$ is a continuous loss function on $\Theta \times U$. 
The following lemma is essential to our result.

\begin{lemma}
\label{lemma:myown}
For a given quantum statistical decision problem, the following statements hold.
\begin{enumerate}[(i)]
\item For any POVM $\Mre \in \povm $, $R_{\Mre}\in C(\Theta )$.
\item A map $\Mre \in \povm \mapsto R_{\Mre} \in C(\Theta )$ is continuous.
\item A set $ \{ R_{\Mre} \in C(\Theta ) :\ \Mre \in \povm \}$ is a compact subset of $ C(\Theta ) $.\\
\end{enumerate}
\end{lemma}

Note that we abandon the assumption of the discrete topology on $\Theta $, which was adopted in previous works~\cite{LeCam1964,Holevo1976}.
Thus, the first assertion is no longer trivial.
Due to the above lemma, we easily show the following theorem. \\


\begin{theorem}\label{theo:risk}
For every $ f \in C(\Theta )$, the following assertions are equivalent:
\begin{enumerate}[(i)]
\item 
There exists $\Mre \in \povm $ such that $f (\theta) \geq R_{\Mre} (\theta )$ for every $\theta \in \Theta $. 
\item 
 $\int f (\theta) \pi (\dtheta )  \geq \inf_{\Mre \in \povm  } \int R_{\Mre}(\theta ) \pi (\dtheta )$
 for every $\pi \in \cal{P} (\Theta ) $.
\end{enumerate}
\end{theorem}

\proof
Obviously (i) implies (ii).
To prove (ii) $\to $ (i), we define
\begin{align*}
& \cal{M}_{1}:= \{ f \} \text{ and} \\
& \cal{M}_{2}:=  \{ R_{\Mre} \in C(\Theta ) :\ \Mre \in \povm \}. 
\end{align*}
Note that $\cal{M}_{2}$ is convex due to the convexity of $\povm$.
We show condition (ii) in Proposition~\ref{pro:LeCam} for $\model_{1}$ and $\model_{2}$.
For every $\pi \in \ptheta $
\begin{align*}
\psi_{\cal{M}_{2}} (\pi ) &= \inf_{g \in \cal{M}_{2}}  \int g (\theta ) \pi (\dtheta ) \\
 &= \inf_{\Mre \in \povm } \int R_{\Mre } (\theta ) \pi (\dtheta ) \\
 & \leq \int f(\theta ) \pi (\dtheta ) \\  
 &= \psi_{\cal{M}_{1}} (\pi )
\end{align*}
holds.
Thus, condition (ii) in Proposition~\ref{pro:LeCam} holds, which implies condition (i) in Proposition~\ref{pro:LeCam}.

Then, due to condition (i) in Proposition~\ref{pro:LeCam}, there exists $g \in \overline{\alpha (\cal{M}_{2})}$
such that $g(\theta ) \leq f(\theta ), \forall \theta \in \Theta$.
In particular $\overline{\alpha (\cal{M}_{2})} = \alpha (\cal{M}_{2}) $ holds from Lemma~\ref{lem:1}.
By definition of $\alpha (\cal{M}_{2})$, there exists $R_{\Mre} (\theta) \in \cal{M}_{2}$
 such that $R_{\Mre} (\theta ) \leq g(\theta ), \forall \theta \in \Theta $.
Thus, we obtain
\[
 f (\theta ) \geq R_{\Mre} (\theta ),\ \forall \theta \in \Theta.
\]
\qed

\begin{theorem}\label{theo:minimax}
For every bounded continuous loss function $w$
\begin{equation} \label{eq:main}
\inf_{\Mre \in \povm } \sup_{\theta \in \Theta}
 R_{\Mre} (\theta )
 = \sup_{\pi \in \ptheta } \inf_{\Mre \in \povm} \int R_{\Mre } (\theta ) \pi (\dtheta ).
\end{equation}
\end{theorem}

\proof
By definition, (l.h.s.) $\geq $ (r.h.s.) in~\eqref{eq:main} holds obviously.
Thus we show the opposite inequality.
We set 
\[
V := \sup_{\pi} \inf_{\Mre} \int R_{\Mre} (\theta ) \pi (\dtheta ).
\]
Since $V < \infty$, we consider $V$ as a constant function on $\Theta $.
Then, clearly
\[
V \geq \inf_{\Mre \in \povm} \int R_{\Mre } (\theta ) \pi (\dtheta ), \ \forall \pi \in \ptheta
\]
holds. 
In the lefthand-side, we fix $\pi \in \ptheta $ and
 we rewrite $V = \int V \pi (\dtheta )$.
Thus, assertion (ii) in Theorem~\ref{theo:risk} is satisfied.
Assertion (i) in Theorem~\ref{theo:risk} implies that there exists a POVM $\Mre_{*} \in \povm $
 such that
\[
V \geq R_{\Mre_{*}} (\theta ) ,\ \forall \theta \in \Theta.
\]
It implies the inequality,
\begin{align*}
V &\geq \sup_{\theta \in \Theta}  R_{\Mre_{*}} (\theta ) \\
  &\geq  \inf_{\Mre \in \povm} \sup_{\theta \in \Theta}  R_{\Mre } (\theta ).
\end{align*}
\qed

Until now, we assume continuous loss functions.
Finally we remove this assumption.
Let $\overline{\real}$ denote the extended real, i.e., $\real \cup \{ + \infty\}$.
We assume that the loss function $w: \Theta \times U \to \overline{\real}$ is lower semicontinuous.
One of the important example is the quantum relative entropy~\cite{Ohya}.
\[
 D(\rho || \sigma ) := 
  \begin{cases} \Tr (\rho \log \rho - \rho \log \sigma ) < \infty,\ \ran \rho \subseteq \ran \sigma \\
				 +\infty,\ \text{otherwise}
  \end{cases}				
\]
with $ \Theta = U = \state$ when $\dim \Hilbert < \infty$.

\begin{theorem}\label{theo:minimax2}
The assertion of Theorem \ref{theo:minimax} is still valid if the loss function $w$ is
 an arbitrary lower semicontinuous $\overline{\real}$-valued function.
\end{theorem}

\proof
Recall that any lower semicontinuous function on a locally compact space is represented by a supremum of 
bounded continuous functions (See, e.g., Bourbaki, Chap.4, section 1.1~\cite{Bourbaki2000} for proof.).
Let us define a class of nonnegative bounded continuous loss functions, 
$ \cal{V} :=\{  v \in C(\Theta \times U):\ 0 \leq v (\theta, u) \leq w(\theta, u), \ \forall \theta, \forall u \} $.
Then for every $\theta$ and $u$ fixed, $w(\theta, u) = \sup \{ v(\theta, u):\ v \in \cal{V} \}$ holds.
In addition
\[
 \int_{U} w (\theta, u) \mu (\rmd u) = \sup_{v \in \cal{V}} \int_{U} v(\theta, u) \mu (\rmd u)
\]
holds for every $\theta \in \Theta$ and every $\mu \in \ptheta $. (See, e.g., Bourbaki, Chap.4, section 1.1~\cite{Bourbaki2000} for the equality.)

We show the inequality (l.h.s.) $\leq $ (r.h.s.) in Eq.~\eqref{eq:main}.
We set
\[
V := \sup_{\pi \in \ptheta } \inf_{\Mre \in \povm} \int R_{\Mre} (\theta ) \pi (\dtheta ).
\]
When $V = \infty$, the desired inequality is satisfied.
Thus, we show the inequality when $V < \infty $.
Then clearly 
\[
 +\infty > \inf_{\Mre \in \povm} \int R_{\Mre} (\theta )\pi (\dtheta ),\ \forall \pi  
\]
holds.
For every nonnegative bounded continuous loss function $v \in \cal{V}$, we have
\begin{align*}
V = \int V \pi (\dtheta ) & \geq \inf_{\Mre \in \povm} \int R_{\Mre} (\theta )\pi (\dtheta ) \\
	& \geq \inf_{\Mre \in \povm} \int R^{v}_{\Mre} (\theta )\pi (\dtheta ),
\end{align*}
where  $R^{v}_{\Mre}$ is the risk function for $v\in \cal{V}$.
From Theorem~\ref{theo:risk}, we may choose a POVM $ \Mre^{(v)} \in \povm$ for the loss $v$ such that
\begin{align*} 
V \geq R^{v}_{\Mre^{(v)} } (\theta ),\ \forall \theta \in \Theta 
\end{align*}
holds.
Now let us consider a subset of $\povm $
\[
\povm^{(v)} := \{ \Mre \in \povm :\ V \geq R_{\Mre }^{v} (\theta) , \forall \theta \in \Theta   \}.
\]
From the above argument, clearly $\povm^{(v)} \neq \emptyset $.
Due to the compactness of $\povm $, we can easily show 
\[
 \bigcap_{v \in \cal{V}} \povm^{(v)} \neq \emptyset.
\]

Finally we choose any element $\DS \Mre_{*} \in \bigcap_{v \in \cal{V}} \povm^{(v)}  $.
For every $\theta \in \Theta $ fixed, 
\[ 
V \geq R_{\Mre_{*} }^{(v)} (\theta ),\ \forall v \in \cal{V} 
\]
holds.
Thus, for any $\theta \in \Theta$, 
\begin{align*}
V &
 \geq \sup_{v \in \cal{V}  } R_{\Mre_{*} }^{(v)} (\theta ) \\
 &= \sup_{v \in \cal{V}  } \int_{U} v(\theta, u) \Tr \rho (\theta) M_{*} (\rmd u) \\
 &= \int_{U} w(\theta, u) \Tr \rho (\theta ) M_{*} (\rmd u) \\
 &= R_{\Mre_{*} } (\theta )
\end{align*}
holds. We obtain 
\begin{align*}
V &\geq \sup_{\theta \in \Theta} R_{\Mre_{*} } (\theta ) \geq 
\inf_{\Mre \in \povm } \sup_{\theta \in \Theta} R_{\Mre} (\theta ).
\end{align*}
\qed

\begin{definition}
We call the value of the equality of Eq.~\eqref{eq:main} the minimax value.
\end{definition}

We obtain the existence theorem of the minimax POVM.
\begin{corollary}
\label{coro:lowersemi}
For every lower semicontinuous $\overline{\real}$-valued loss function, 
 there exists a minimax POVM.
\end{corollary}

\proof
We take a POVM $\Mre_{*}$ in the proof of Theorem~\ref{theo:minimax2}, which is shown to be minimax. 
\qed

As far as the author knows, minimax theorem has not been shown yet in quantum statistical decision theory.
Partial and insufficient result was obtained in Hirota and Ikehara~\cite{Hirota1982}.
However, their result seems to be an immediate consequence of result by Wald~\cite{Wald1950} because their statement was restricted to finite-dimensional quantum systems.

Existence of minimax POVM itself was shown by Bogomolov~\cite{Bogomolov1982} under weaker assumptions than ours.
However, as is known in classical statistics, minimax theorem~\ref{theo:minimax2} does not hold necessarily in their broad assumptions.
For counterexample, see, e.g., Ferguson~\cite{Ferguson}, section 2.9, p.83.

Finally, we emphasize that our result is not only of theoretical interest but also of practical significance.
\begin{corollary}
\label{coro:minimax}
Theorem \ref{theo:minimax} and Theorem \ref{theo:minimax2} and corollaries derived from them 
are valid if we take a closed convex subset of $\povm $ instead of all POVMs $\povm$.
\end{corollary}

For example, in a composite system $\Hilbert_{1} \otimes \Hilbert_{2}$, 
 we might consider all of separable measurements.
Practically, we are able to restrict POVMs to experimentally realizable class of measurements.

\section{Least Favorable Priors}

\subsection{Least favorable priors}

In Bayesian statistics, least favorable priors have some significance~\cite{Bernardo2005}.
\begin{definition}
\label{defi:lfp}
A distribution $\pi_{*} \in \ptheta $ is called a least favorable prior if it satisfies
\[
\inf_{\Mre \in \povm} \int R_{\Mre } (\theta ) \pi_{*} (\dtheta )
 = \sup_{\pi \in  \ptheta } \inf_{\Mre \in \povm} \int R_{\Mre } (\theta ) \pi (\dtheta ).
\]
\end{definition}

We obtain the existence theorem of least favorable priors.

\begin{theorem}
\label{theo:lfp}
For every continuous loss function,  there exists a least favorable prior $\pi_{LF} \in  \ptheta $.
Every minimax POVM is a Bayesian POVM with respect to $\pi_{LF}$.
\end{theorem}

\proof
Let us introduce the weak topology in $ \ptheta $. 
It is well known that $ \ptheta $ is compact when $\Theta $ is compact.

By Lemma~\ref{lemma:myown}, $R_{\Mre} \in C(\Theta)$ holds and a function 
\[
 \pi \mapsto I_{\Mre } (\pi ) := \int_{\Theta} R_{\Mre } (\theta ) \pi (d \theta ), \ \pi \in \ptheta
\]
is clearly affine and continuous with respect to the weak topology.
Thus,
\[
 I_{\povm } (\pi ) = \inf_{\Mre \in \povm} I_{\Mre} (\pi )
\]
is a upper semicontinuous function and achieves the maximum, i.e., there exists $\pi_{LF} \in \ptheta $
 such that
\[ 
 I_{\povm} (\pi_{LF}) =  \sup_{\pi } I_{\povm } (\pi ),
\]
 which implies $\pi_{LF}$ is a least favorable prior.

For the latter part, we define
\[
V := \inf_{\Mre \in \povm } \sup_{\theta \in \Theta} R_{\Mre} (\theta ).
\]
If $V = \infty$, for every POVM $\Mre \in \povm$,
\[
\int R_{\Mre} (\theta ) \pi_{LF}(d \theta ) = \infty
\]
holds from Theorem~\ref{theo:minimax2}
and thus every POVM is Bayesian with respect to $\pi_{LF}$.
(Likewise, every POVM is also minimax.)

When $V < \infty$, we take an arbitrary minimax POVM $\Mre_{*}$.
Clearly 
\[
 V \geq R_{\Mre_{*}} (\theta ),\ \forall \theta \in \Theta  
\]
holds.
Thus, taking average with respect to $\pi_{LF}$, 
\begin{align*}
V \geq \int R_{\Mre_{*}} (\theta ) \pi_{LF}(d \theta).
\end{align*}
On the other hand,  from Theorem~\ref{theo:minimax2}, 
\[
V = \inf_{\Mre} \int R_{\Mre} (\theta ) \pi_{LF} (d\theta )
\]
is the smallest average risk with respect to $\pi_{LF}$, which implies $\Mre_{*}$
 is a Bayesian POVM with respect to $\pi_{LF}$.
\qed



\begin{remark}
Historically speaking, Holevo~\cite{Holevo1976} first showed the existence of the Bayes POVM and 
Ozawa~\cite{Ozawa1980} also showed the same result in a broader context.
Thus, in our setting, there exists a Bayes POVM with respect to $\pi_{LF}$.
One immediate consequence is the following corollary.
\end{remark}

\begin{corollary}
\label{coro:lfp2}
For every continuous loss function, every minimax POVM is among the class of Bayes POVMs with respect to
 a least favorable prior.
If a Bayes POVM with a least favorable prior is unique, then it is minimax.
\end{corollary}

Generally the uniqueness of the Bayes POVM depends on a loss function, but at least in classical statistics 
 the uniqueness is shown for typical loss functions.

Some readers might wonder whether the above assertion holds or not if we adopt a lower semicontinuous function 
as the loss function.
Unfortunately, when a bounded lower semicontinuous (and not continuous) loss function is adopted,
a statistical decision problem may have no least favorable prior.
Let us consider the following pathological loss that is independent of decision.

\begin{example}
Pathological loss.
Let $\Theta =[0,1] \subseteq \real$.
We consider a sequence of continuous loss function, $L^{n} (\theta, u)$, which is defined by
\begin{align*}
R^{n}(\theta ) = L^{n}(\theta, u) = \begin{cases}
				1- \theta, \ 1/n \leq \theta \leq 1, \\
				(n-1) \theta, \ 0 \leq \theta \leq 1/n.
 \end{cases}
\end{align*}
Since the loss is independent of decision $u$, it is equal to the risk function for any decision.
The supremum of the sequence $\{ R^{n}\}$ is no longer a continuous function but a bounded lower-semicontinuous function,
\begin{align*}
R(\theta ) = \sup_{n} R^{n} (\theta )  = \begin{cases}
				1- \theta, \ 0 < \theta \leq 1, \\
				0, \ \theta =0.  
 \end{cases}
\end{align*}
For this risk function, any POVM is minimax and Bayes.
The minimax value is $V = \sup_{\theta \in [0, 1]} R(\theta ) =1$.
If a least favorable prior $\pi_{LF}$ would exist, then it would satisfy
\[
 1 = \int R(\theta ) \pi_{LF} (d \theta ),
\]
 which is clearly impossible.
In particular,  Lemma~\ref{lemma:myown} (i) does not hold any more.
In spite of this, quantum minimax theorem (Theorem~\ref{theo:minimax2}) still holds.

We also see that a sequence of least favorable prior have a limit point in terms of the weak topology
due to the compactness of $\Theta = [0,1]$ and the limit distribution is a prior but not a least favorable prior. 
For each $n$, the least favorable prior for $L^{n}$ is obviously given by
$\pi_{LF}^{n} = \delta_{1/n}$, where $\delta_{a} $ denotes the Dirac distribution (i.e., the point $\theta =a$ has the whole mass one).
Its weak limit is given by $ \pi^{\infty} = \delta_{0}$.
Clearly, 
\[
 V > \int R(\theta ) \pi^{\infty} (d \theta ) = R(0) =0.
\]

\end{example}

\begin{remark}
For lower-semicontinuous loss function, the assertions in Theorem~\ref{theo:lfp} and Corollary~\ref{coro:lfp2} 
 hold if a least favorable prior exists.
\end{remark}

As in the last section, we emphasize that our result in this section holds when we restrict POVMs to a smaller class.
\begin{corollary}
\label{coro:lfp}
Theorem~\ref{theo:lfp} and Corollary~\ref{coro:lfp2} are valid if we take a closed convex subset of $\povm $ instead of all POVMs $\povm$.
\end{corollary}

\subsection{Connection with some results in Bayesian statistics}



The classical statistical decision problem is regarded as a special case of quantum statistical decision problem. 
Thus our result would apply to classical statistics.
Rigorous treatment will be presented in another occasion.
Here, we take an example and give a rough idea of this assertion.

A least favorable prior is called a latent information prior~\cite{Komaki2011} when we consider statistical predictions and adopt the Kullback divergence as a loss function, which is a $\overline{\real}$-valued lower semicontinuous function.
In this specific loss function, the compactness of $ \ptheta $ assures the existence of least favorable priors while Theorem 4 does not apply to this case.
However, the minimax theorem would apply to the problem and thus, the Bayesian decision, which is called Bayesian predictive distribution in this context, is shown to be unique and minimax (section 3 in Komaki~\cite{Komaki2011}).
In the same way, recent results and even old ones in mathematical statistics would be partially obtained as a consequence of our result.

\section{Concluding remarks}
\label{Conclu}

In conclusion, we emphasize that our theorem holds for every closed convex subset of POVMs.
The classical analogue of such a restricted class of POVMs is a restricted class of estimators,
 which seems unusual in theoretical analysis.
Thus, this concept becomes much more meaningful in quantum statistical decision theory
 than in classical one.
Also, it is important in practice to find an optimal POVM among a restricted class, which depends on each physical setup and experimenter's preference.
Further investigations in this direction, e.g., numerical algorithm to find least favorable priors as in classical Bayesian statistics
 are left for future studies but surely go beyond a straightforward extension of classical results.


\appendix

\section{Proof of Lemma~\ref{lemma:myown}}
\label{proofs}


\subsection{Proof of Lemma~\ref{lemma:myown} (i)}
%
%

\proof
Let $\Mre $ be a POVM, $\rho (\theta )$ be a quantum statistical model and $f$ be a nonnegative continuous function on a compact space $\Theta \times U$.
Let $C$ be a constant such that $ f (\theta, u) \leq C$ for all $\theta \in \Theta$ and all $u \in U$. 
For each $\theta, \theta' \in \Theta$, we have
\begin{align*}
| R_{\Mre}(\theta ) - R_{\Mre} (\theta' ) | & 
\leq 
 \left| \int_{U} f(\theta, u) \Tr \rho (\theta ) M(\rmd u) -  \int_{U} f(\theta', u) \Tr \rho( \theta') M(\rmd u)  \right| \\
 & \leq 
   \left| \int_{U} ( f(\theta, u)- f(\theta', u) ) \Tr \rho (\theta ) M(\rmd u) \right| \\
 & \qquad+ \left| \int_{U} f(\theta', u) \Tr \{ \rho( \theta) - \rho (\theta') \} M(\rmd u)  \right|. 
\end{align*}

First we evaluate the second term in the last inequality.
Since $ \rho (\theta ) - \rho (\theta') \leq | \rho (\theta ) - \rho (\theta') | $ and $f \geq 0$, we have 
\begin{align*} 
\left| \int_{U} f(\theta', u) \Tr \{ \rho ( \theta ) - \rho ( \theta' ) \}  M(\rmd u)  \right|
&\leq  \int_{U}  f(\theta', u) \Tr | \rho (\theta)  - \rho (\theta') | M(\rmd u)   \\
&\leq  C  \int_{U} \Tr |\rho ( \theta ) - \rho ( \theta' )| M(\rmd u)   \\
 &= C\Tr |\rho (\theta ) - \rho ( \theta' ) |.
\end{align*}
Due to the regularity, the second term in the last equality, obviously goes to $0$ as $d (\theta, \theta') \migi 0$.
For the first term, given $\theta \in \Theta$, we have
\begin{align*}
\lim_{\theta' \migi \theta } \int_{U} f(\theta', u)  \Tr \rho (\theta ) M(\rmd u) 
 = \int_{U} f(\theta, u)  \Tr \rho (\theta ) M(\rmd u).
\end{align*}
%
Thus, at each point $\theta \in \Theta$, $R_{\Mre} (\theta )$ is continuous. \qed

\subsection{Proof of Lemma~\ref{lemma:myown} (ii) and (iii)}

From Lemma~\ref{lemma:myown} (i), we may consider $R_{\Mre} (\theta)$ as a $C(\Theta )$-valued function over the set of POVMs.
\begin{align*} 
\Mre \in \povm \mapsto R_{\Mre} (\cdot ) \in C(\Theta ).
\end{align*}

Since the continuous image of a compact set is compact, the assertion (iii) in Lemma~\ref{lemma:myown} directly follows from the assertion (ii).
Thus, it is enough to show the assertion (ii) in Lemma~\ref{lemma:myown}.
In order to make the proof concise, we first show the following lemmas.

\begin{lemma}
\label{lem:bound}
Let $\Theta $ be a compact metric space and $\Hilbert$ be a Hilbert space.
If a function $T$ from $\Theta $ to $L_{1}(\Theta )$ is regular, then there is a trace-class operator $T_{\Theta}$ satisfying
\begin{equation} \label{unifbound} 
 T(\theta ) \leq T_{\Theta},\ \ \forall \theta \in \Theta 
\end{equation}
\end{lemma}

\proof
Without loss of generality, we assume $T (\theta ) \geq 0 $ for every $\theta \in \Theta$.
We choose $\delta > 0$ such that $\omega_{T} (\Theta_{\delta}) < \infty$.
Then, there is a trace-class operator $X$ satisfying
\[
 -X \leq T (\theta ) - T (\eta )  \leq X, 
\]
for each $(\theta, \eta) \in \Theta_{\delta}$.
Due to compactness of $\Theta$, there exist finite points $\theta_{1}, \dots, \theta_{m}$ such that
for every $\eta \in \Theta$, $d(\eta, \theta_{j}) < \delta$ for some $j$, $1 \leq j \leq m$.

Now we take $T_{\Theta}$ as
\[
T_{\Theta} := X + \sum_{i=1}^{m} T( \theta_{i}).
\]
Clearly $T_{\Theta }$ satisfies the inequality~\eqref{unifbound}. \qed

\begin{remark}
Even if $\sup\{ \Tr|T(\theta )|:\ \forall \theta \in \Theta \} < \infty$ holds, a trace-class operator 
satisfying \eqref{unifbound} does not necessarily exist. 
\end{remark}

\begin{lemma}\label{equiregular}
Let $\rho$ be a quantum statistical model and $f$ a nonnegative continuous function on a compact set $\Theta \times U$.
Set $\kappa_{u} ( \theta ) := f(\theta, u) \rho (\theta) $.
Then, given $\epsilon > 0$ and a compact set $K \subseteq \Theta $, there is a trace-class operator $W$ and $\delta > 0$ such that 
\begin{align}
 & -W \leq \kappa_{u} (\theta )  - \kappa_{u} (\eta ) \leq W, \ \forall (\theta, \eta) \in K_{\delta}, \forall u \in U, \nonumber \\
 &\Tr W \leq \epsilon. \label{cond:w}
\end{align}
\end{lemma}

\proof
It is enough to show the assertion when $K=\Theta$.

First, let us introduce
\begin{align*}
 & \omega_{f}(\Theta_{\delta}, U) := \sup \{ |f(\theta_{1}, u) - f(\theta_{2}, u)|:\  (\theta_{1}, \theta_{2}) \in \Theta_{\delta }, u \in U \}. 
\end{align*}
It is easily seen that $ \omega_{f}(\Theta_{\delta}, U) \migi 0 $ as $\delta \migi 0$. 
Thus, we take $\delta $ such that 
\[
 \omega_{f}(\Theta_{\delta}, U) \leq \epsilon,\  \omega_{\rho}(\Theta_{\delta}) \leq \epsilon.
\]

Now, due to the regularity of $\rho$, there is a trace-class operator $X$ satisfying
\begin{align*}
 & -X \leq \rho (\theta ) - \rho (\eta)  \leq X,\ \forall (\theta, \eta ) \in \Theta_{\delta }, \\
 & \Tr |X| \leq  2 \epsilon. 
\end{align*}
Then, for every $(\theta, \eta) \in \Theta_{\delta} $ and  for every $u \in U$, 
\[
|f(\theta, u)| \{ \rho (\theta) - \rho (\eta ) \} \leq C X,
\]
 where $C := \sup \{ |f(\theta, u)| :\ \theta \in \Theta, u \in U  \}$.

Due to Lemma~\ref{lem:bound}, there is a trace-class operator $Z$ satisfying
\begin{align*}
 -Z \leq \rho ( \theta ) \leq Z, \ \forall \theta \in \Theta.
\end{align*}



Thus, for every $(\theta, \eta) \in \Theta_{\delta}$ and $u \in U$, we have
\begin{align*}
 \kappa_{u} (\theta ) - \kappa_{u}( \eta ) 
& = \{ f(\theta, u) - f(\eta, u) \} \rho ( \theta )  +f(\eta, u) \{ \rho (\theta ) - \rho (\eta ) \}  \\
& \leq   \omega_{f}(\Theta_{\delta}, U) Z + C X \\
& \leq \epsilon Z + C X
\end{align*}
and 
\begin{align*}
 \Tr (\epsilon Z + C X) 
 & \leq \epsilon (\Tr Z + 2C).
\end{align*}

If we replace $\epsilon$ with $ \epsilon/ (\Tr Z + 2C) $ and repeat from the beginning, we obtain  a trace-class operator $W$ satisfying \eqref{cond:w}.
\qed


We review some definitions for reader's convenience.
\begin{definition}
A family $\cal{F}$ of functions from a metric space $X$ to a metric space $Y$ with metric $d$
  is called equicontinuous at the point $x \in X$ if given $\epsilon > 0$ there is a open set $O_{x}$ containing $x$
 such that $d( f(x), f(x') ) < \epsilon $ for all $x'$ in $O_{x}$ and for all $f $ in $\cal{F}$.
The family is said to be equicontinuous on $X$ if it is equicontinuous at each point $x$ in $X$.
\end{definition}

Now, we show the equicontinuity.
\begin{lemma}
A family of continuous functions, $\cal{F}:= \{ R_{\Mre} \in C(\Theta ):\ \Mre \in \povm \}$, is (uniformly) equicontinuous on $\Theta $.
\end{lemma}

\proof
From Lemma \ref{equiregular}, for every $\epsilon > 0$, 
there is a trace-class operator $W$ and $\delta > 0$ satisfying \eqref{cond:w}.
For every measurement $\Mre \in \povm$ and every  $ (\theta_{1}, \theta_{2}) \in \Theta_{\delta}$, we have
\begin{align*}
| R_{\Mre} (\theta_{1}) - R_{\Mre } (\theta_{2} ) | & \leq \left| \int_{U}  \Tr \kappa_{u}(\theta_{1}) M(\rmd u ) -\int_{U}  \Tr \kappa_{u} (\theta_{2} ) M(\rmd u )  \right| \\
& \leq \int_{U} \Tr W M(\rmd u) \\
& \leq \epsilon.
\end{align*}
\qed

Finally, we show Lemma~\ref{lemma:myown} (ii).

\proof
Let $\Mre_{0} \in \povm $ and $\epsilon > 0$ fixed.
We construct a neighborhood of $\Mre_{0} $, denoted as $\cal{W}(\Mre_{0}) \subseteq \povm$ and show that
it satisfies
\[
 \Mre \in \cal{W}(\Mre_{0}) \Rightarrow 
 \norm{ R_{\Mre_{0}} - R_{\Mre} }:= \sup_{\theta \in \Theta} |R_{\Mre_{0}}(\theta ) - R_{\Mre} (\theta ) | \leq \epsilon. 
\]

From the equicontinuity at each point $\theta \in \Theta $, we may choose $\delta_{\theta} > 0$ such that
\[
 d(\theta, \eta) < \delta_{\theta} \Rightarrow 
  | R_{\Mre} (\theta ) - R_{\Mre } (\eta  ) | < \frac{\epsilon}{3}, \forall \Mre \in \povm.
\]
Now we define an open neighborhood of $\theta$ as
\[
 V_{\theta} := \{ \eta \in \Theta: \  d(\theta, \eta ) < \delta_{\theta} \}.
\]
Due to compactness of $\Theta $, there exists a finite number of open sets $V_{\theta_{1}}, \dots, V_{\theta_{m}}$ 
 satisfying $\Theta = V_{\theta_{1}} \cup \dots \cup V_{\theta_{m}}$.
For each point $\theta_{1}, \dots, \theta_{m}$, we take an open neighborhood defined by  
\begin{align*}
\cal{U}_{j} 
&:= \{ \Mre \in \povm :\ | R_{\Mre_{0}}( \theta_{j}) - R_{\Mre}( \theta_{j} )  | < \epsilon /3  \}.
\end{align*}
Then we set
\begin{align*}
\cal{W} (\Mre_{0}) := \bigcap_{j=1}^{m} \cal{U}_{j}.
\end{align*}

For every POVM $\Mre \in \cal{W}(\Mre_{0})$ and every $\eta \in \Theta $, there exists $\theta_{j}$ satisfying $d (\theta_{j}, \eta) < \delta_{\theta_{j}} $.
Thus we have
\begin{align*}
| R_{\Mre_{0}}( \eta ) - R_{\Mre}( \eta  )  | & \leq | R_{\Mre_{0}}( \eta ) - R_{\Mre_{0}}( \theta_{j}) | +| R_{\Mre_{0}}( \theta_{j}) - R_{\Mre} (\theta_{j} )  |  + | R_{\Mre}( \theta_{j}) - R_{\Mre} (\eta )  |  \\
 & < \epsilon, 
\end{align*}
where the difference $ |R_{\Mre_{0}}( \theta_{j}) - R_{\Mre} (\theta_{j} )  |$ is bounded because $ \Mre \in \cal{W}(\Mre_{0})$.
The other two terms are bounded due to the equicontinuity.
Thus, for every POVM $\Mre \in \cal{W}(\Mre_{0})$, we obtain
\[ 
\norm{ R_{\Mre_{0}} - R_{\Mre} } \leq \epsilon.
\]
It implies that the map from $\povm $ to $C(\Theta)$ is continuous. \qed

%
%

\section*{Acknowledgement}
This work was supported by the Grant-in-Aid for Young Scientists (B) (No. 24700273) and the Grant-in-Aid for Scientific Research (B) (No. 26280005).
The author is grateful to Dr. Sakashita and Dr. Sugiyama for fruitful discussions.

%
%

\end{document}